\documentclass{elsart}
\usepackage{amsmath}
\usepackage{graphicx}

\begin{document}
\begin{frontmatter}
\title{A microscopic model of triangular arbitrage}
\author{Yukihiro Aiba$^{\rm a, 1}$, Naomichi Hatano$^{\rm b}$}
 
\thanks{e-mail: aiba@iis.u-tokyo.ac.jp}
\address{
$^{\rm a}$Department of Physics, University of Tokyo\\
Hongo 7-3-1, Bunkyo, Tokyo, Japan}
\address{
$^{\rm b}$Institute of Industrial Science, University of Tokyo\\
Komaba 4-6-1, Meguro, Tokyo, Japan}

\begin{keyword}
Econophysics;
Triangular Arbitrage:
Financial Market;
Foreign Exchange;
Agent Model
\PACS{05.40.-a; 05.90+m; 89.65.-s; 89.65.Gh}
\end{keyword}

\begin{abstract}
We introduce a microscopic model which describes the dynamics of each dealer in multiple  foreign exchange markets, taking account of the triangular arbitrage transaction.
The model reproduces the interaction among the markets well. 
We explore the relation between the parameters of the present microscopic model and the spring constant of a macroscopic model that we proposed previously.
\end{abstract}
\end{frontmatter}

\section{Introduction}
Analyzing correlation in financial time series is a topic of considerable interest \cite{6:mandelbrot,6:fama,6:ding,6:dacorogna,6:liu,6:mantegnabook,6:mantegna2,6:plerou,6:bouchaud,6:Plerou,6:kullmann,6:mandelbrot2,6:kullmann2,6:onnela,6:mizuno,6:tastan,6:toth,6:jung}.
We recently pointed out \cite{6:tat1,6:acf,6:tat3} that, in the foreign exchange market, a correlation among exchange rates can be generated by triangular arbitrage transactions.
The triangular arbitrage transaction is a financial activity that takes advantage of the three foreign exchange rates among three currencies \cite{6:tat1,6:book,6:moosa}. 
It makes the product of the three foreign exchange rates converge to its average, thereby generating an interaction among the rates.

In order to study effects of the triangular arbitrage on the fluctuations of the exchange rates, we introduced \cite{6:tat1} a stochastic model describing the time evolution of the exchange rates with an interaction. The model successfully described the fluctuation of the data of the real market.
The model is phenomenological; {\it i.e.} it treats the fluctuations of the rates as fluctuating particles and the interaction among the rates as a spring. 
We refer to this model as the \lq macroscopic model\rq\ hereafter.

The purpose of this paper is to understand {\it microscopically} effects of the triangular arbitrage on the foreign exchange market.
For the purpose, we introduce a new model which focuses on each dealer in the markets; we refer to the new model as the \lq microscopic model\rq\ hereafter. 
We then show the relation between  the macroscopic model and the microscopic model through an interaction strength which is regarded as a spring constant.

The paper is organized as follows.
We first review the macroscopic model of the triangular arbitrage in Section \ref{6:macromodel}. 
Second, in Section \ref{6:micromodel}, we introduce the microscopic model which focuses on the dynamics of each dealer in the markets. 
The model reproduces the interactions among the markets well. 
We explore the relation between the spring constant of the macroscopic model and the parameters in the microscopic model in Section \ref{6:micromacro}.
We summarize the paper in Section \ref{6:sum}.

\section{Review of macroscopic model of triangular arbitrage}
\label{6:macromodel}
The triangular arbitrage is a financial activity that takes advantage of three exchange rates.
When a trader exchanges one Japanese yen to some amount of US dollar, exchanges the amount of US dollar to some amount of euro and exchanges the amount of euro back to Japanese yen instantly at time $t$, the final amount of Japanese yen is given by
\begin{equation}
\mu\equiv\prod_{x=1}^{3}r_x(t),
\end{equation}
where
\begin{align}
r_1(t)&\equiv\ \frac{1}{\mbox{yen-dollar ask }(t)}\label{6:eq:defr1}\\
r_2(t)&\equiv\ \frac{1}{\mbox{dollar-euro ask }(t)}\label{6:eq:defr2}\\
r_3(t)&\equiv\ \mbox{yen-euro bid }(t).\label{6:eq:defr3}
\end{align}
Here, \lq bid\rq\ and \lq ask,\rq\ respectively, represent the best bidding prices to buy and to sell in each market.
If the rate product $\mu$ is greater than unity, the trader can make profit through the above transaction.
This is the triangular arbitrage transaction.
Once there is a triangular arbitrage opportunity, many traders will make the transaction.
This makes $\mu$ converge to a value less than unity, thereby eliminating the opportunity.
Triangular arbitrage opportunities nevertheless appear, because each rate $r_i$ fluctuates strongly. 

\begin{figure}
 \begin{minipage}{6cm}
 \begin{center}
 \includegraphics{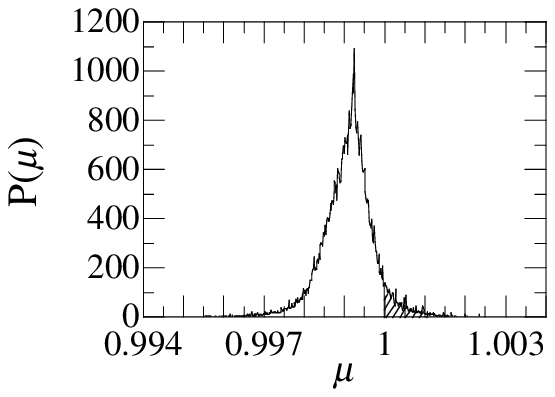}\\
 (a)
 \end{center}
 \end{minipage}
 \quad
 \begin{minipage}{6cm}
 \begin{center}
 \includegraphics{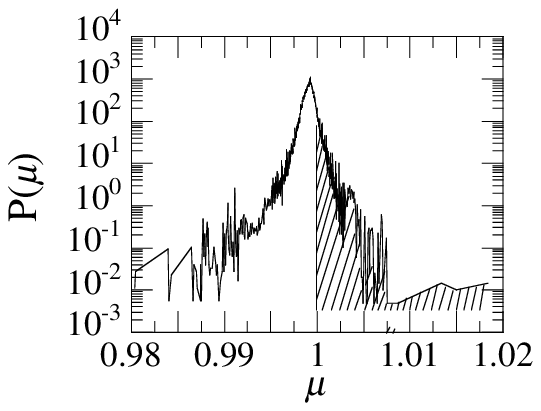}\\
 (b)
 \end{center}
 \end{minipage}
  \caption{The probability density function of the rate product $\mu$ \cite{6:tat1}. (b) is a semi-logarithmic plot of (a). The shaded area represents triangular arbitrage opportunities. The data were taken from January 25 1999 to  March 12 1999.}
  \label{6:fig:fils}
\end{figure}

The probability density function of the rate product $\mu$ has a sharp peak and  fat tails (Fig.\ \ref{6:fig:fils}).
It means that the fluctuations of the exchange rates have correlation that makes the rate product converge to its average $\langle\mu\rangle \simeq 0.99998$.
The average is less than unity because of the spread; the spread is the difference between the ask and the bid prices and is usually of the order of $0.05\%$ of the price.

For later convenience, we here define the logarithm rate product $\nu$ as the logarithm of the product of the three rates:
\begin{equation}
\label{6:eq:defnu}
  \nu(t) = \ln\prod_{x=1}^{3}r_x(t) = \sum_{x=1}^{3}\ln r_x(t).
\end{equation}
There is a triangular arbitrage opportunity whenever this value is positive.

We can define another logarithm rate product $\nu\rq$, which has the opposite direction of the arbitrage transaction to $\nu$, that is, from Japanese yen to euro to US dollar back to Japanese yen:

\begin{equation}
\label{6:eq:defnu'}
  \nu\rq(t) = \sum_{x=1}^{3}\ln {r\rq}_x(t),
\end{equation}
where
\begin{align}
{r\rq}_1(t)&\equiv\ \mbox{yen-dollar bid }(t)\label{6:eq:defr'1}\\
{r\rq}_2(t)&\equiv\ \mbox{dollar-euro bid }(t)\label{6:eq:defr'2}\\
{r\rq}_3(t)&\equiv\ \frac{1}{\mbox{yen-euro ask }(t)}.\label{6:eq:defr'3}
\end{align}
This logarithm rate product $\nu\rq$ will appear in Section \ref{6:micromodel2}.

\begin{figure}
 \begin{minipage}{5.6cm}
 \centering
  \includegraphics{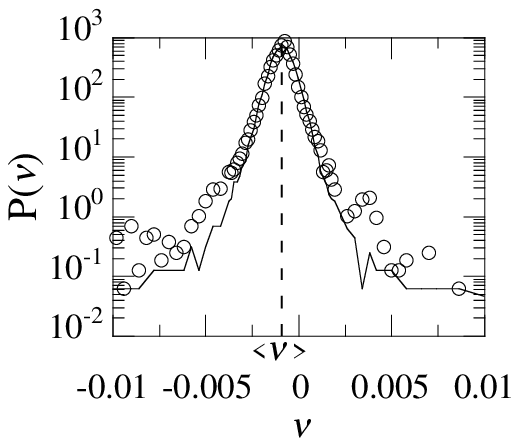} 
  \caption{The probability density function of $\nu$ \cite{6:tat1}. 
    The circle ($\circ$) denotes the real data and the solid line denotes our simulation data. 
	We fix the time step at $T=60$ [sec] and hence use the spring constant $k=0.17$ for our simulation.
    The simulation data fit the real data well.}
  \label{6:fig:compfil}
 \end{minipage}
 \quad
 \begin{minipage}{5.6cm}
 \centering
  \includegraphics[height=4cm]{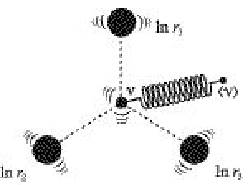}
  \vspace{3.2mm}
  \caption{A schematic image of the model \cite{6:proc}. The three random walkers with the restoring force working on the center of gravity.}
  \label{6:fig:model}
 \end{minipage}
\end{figure}

In one of our previous works\ \cite{6:tat1}, we constructed a stochastic model of the time evolution of foreign exchange rates that takes account of the effect of the triangular arbitrage transaction.
The basic equation of the model is the time evolution of the logarithm of each rate:
\begin{equation}
  \label{6:eq:time.ev.lnr}
  \ln r_x(t+T) = \ln r_x(t)+\eta_x(t)+g(\nu(t)), \ \ \ (x=1,2,3)
\end{equation}
where $T$ is a time step which controls the time scale of the model; we later use the actual financial data every  $T$[sec].
The variable $\eta_x$ denotes an independent fluctuation that obeys a truncated L\'{e}vy distribution\ \cite{6:bouchaud,6:mantegnabook,6:mantegna} and $g$ represents an interaction function defined by
\begin{equation}
\label{6:eq:defg}
g(\nu)=-k(\nu-\langle\nu\rangle),
\end{equation}
where $k$ is a positive constant which specifies the interaction strength and $\langle\nu\rangle$ is the time average of $\nu$.
The time-evolution equation of the logarithm rate product $\nu$ is given by summing Eq.\ (\ref{6:eq:time.ev.lnr}) over all $x$:
\begin{equation}
\label{6:eq:time.ev.nu}
\nu(t+T)-\langle\nu\rangle =(1- 3 k)(\nu(t)-\langle\nu\rangle)+\sum_{x=1}^3 \eta_{x}(t).
\end{equation}
The model equation\ (\ref{6:eq:time.ev.nu}) well describes a fat-tail probability distribution of $\nu(t)$ of the actual market\ (Fig.\ \ref{6:fig:compfil})\ \cite{6:tat1} as well as a negative auto-correlation of the price fluctuation \cite{6:acf}.

From a physical viewpoint, we can regard \cite{6:proc} the model equation (\ref{6:eq:time.ev.lnr}) as a one-dimensional random walk of three particles with a restoring force, by interpreting $\ln r_x$ as the position of each particle (Fig.\ \ref{6:fig:model}). 
The logarithm rate product $\nu$ is the summation of $\ln r_x$, hence is proportional to the center of gravity of the three particles.
The restoring force $g(\nu)$ makes the center of gravity converge to a certain point $\langle\nu\rangle$.
The form of the restoring force (\ref{6:eq:defg}) is the same as that of the harmonic oscillator. 
Hence we can regard the coefficient $k$ as a spring constant.

\begin{figure}
\begin{center}
\includegraphics[height=5cm]{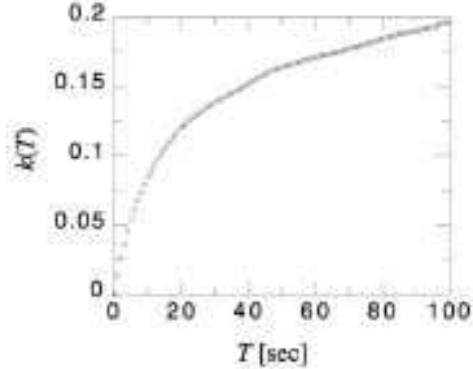}
\caption{The spring constant $k$ as a function of the time step $T$. The spring constant $k$ increases with the time step $T$.}
\label{6:fig:aT}
\end{center}
\end{figure}

The spring constant $k$ is related to the auto-correlation function of $\nu$ as follows \cite{6:tat1}:
\begin{equation}
\label{6:eq:defa}
	1-3 k=c(T)\equiv\frac{\langle \nu(t+T)\nu(t)\rangle - \langle\nu(t)\rangle ^2}{ \langle\nu(t)^2\rangle -\langle\nu(t)\rangle ^2}.
\end{equation}
Using Eq.~(\ref{6:eq:defa}), we can estimate the spring constant $k(T)$ from the real data series as a function of the time step $T$ (Fig.~\ref{6:fig:aT}).
The spring constant $k$ increases with the time step $T$.
We fixed the time step at $T=60$[sec] and hence used the spring constant $k=0.17$ for our simulation.
We will come back to this point later in Section \ref{6:micromacro}.

\section{Microscopic model of triangular arbitrage}
\label{6:micromodel}
We here introduce a microscopic model which describes interactions among foreign exchange markets.
The model focuses on the dynamics of each dealer in the market. 

In order to describe {\it each} foreign exchange market microscopically, we use Sato and Takayasu\rq{s} dealer model (the ST model) \cite{6:sato}, which reproduces the power-law behavior of price changes in a single market well.
Although we focus on the interactions among three currencies, two of the three markets can be regarded as one effective market \cite{6:tat3}; {\it i.e.} the yen-euro rate and the euro-dollar rate are combined to an effective yen-dollar rate. 
In terms of the macroscopic model, we can redefine a variable $r_2$ as the product of $r_2$ and $r_3$.
Then the renormalized variable $r_2$ follows a similar time-evolution equation.
We therefore describe triangular arbitrage opportunities with only two interacting ST models.
\subsection{Sato and Takayasu's dealer model (ST model)}
\label{6:STmodel}
\begin{figure}
\includegraphics[height=4.5cm]{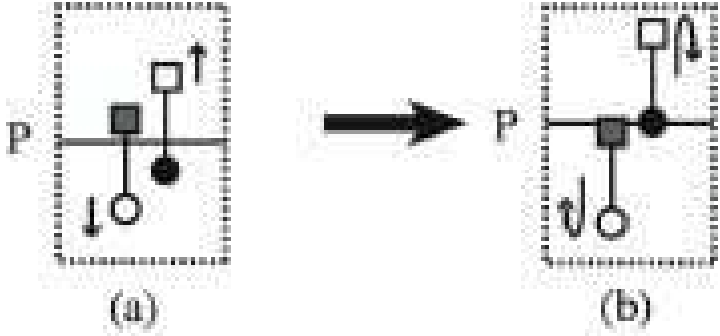}
\caption{A schematic image of a transaction of the ST model.
			Only the best bidders are illustrated in order to simplify the image.
			The circles denote the dealers\rq\ bidding price to buy and the squares denote the dealers\rq\ bidding price to sell.
			The filled circles denote the best bidding price to buy, ${\rm max}\{B_i\}$, and the gray squares denote the best bidding price to sell, ${\rm min}\{B_i\}+\Lambda$.
			In (a), the condition (\ref{6:eq:condition}) is not satisfied, and the dealers, following Eq.~(\ref{6:eq:priceevo}), change their relative positions by $a_i$.
			Note that the term $c\Delta P$ does not depend on $i$; hence it does not change the relative positions of dealers but change the whole dealers\rq\ positions.
			In (b), the best bidders satisfy the condition (\ref{6:eq:condition}).
			The price $P$ is renewed according to Eq.~(\ref{6:eq:price}), and the buyer and the seller, respectively, become a seller and a buyer according to Eq.~(\ref{6:eq:ai}).}
\label{6:fig:intra}
\end{figure}

We first review the ST model briefly \cite{6:sato} (Fig.~\ref{6:fig:intra}).
The basic assumption of the ST model is that dealers want to buy stocks or currencies at a lower price and to sell them at a higher price.
There are $N$ dealers; the $i$th dealer has bidding prices to buy, $B_i(t)$, and to sell, $S_i(t)$, at time $t$.
Let us assume that the difference between the buying price and the selling price is a constant $\Lambda\equiv S_i(t) -B_i(t) > 0$ for all $i$, in order to simplify the model.

The model assumes that a trade takes place between the dealer who proposes the maximum buying price and the one who  proposes the minimum selling price.
A transaction thus takes place when the condition
\begin{equation}
\label{6:eq:condition0}
{\rm max}\{B_i(t)\} \ge {\rm min}\{S_i(t)\}
\end{equation}
or
\begin{equation}
\label{6:eq:condition}
{\rm max}\{B_i(t)\} - {\rm min}\{B_i(t)\} \ge \Lambda
\end{equation}
is satisfied, where max\{$\cdot$\} and min\{$\cdot$\}, respectively, denote the maximum and the minimum values in the set of the dealers\rq\ buying threshold \{$B_i(t)$\}. 
The logarithm of the rates in the actual market, $-\ln r_x$ and $\ln r\rq_x$, correspond to max\{$B_i$\} and min\{$S_i$\}, respectively.

The market price $P(t)$ is defined by the mean value of max\{$B_i$\} and min\{$S_i$\} when the trade takes place. 
The price $P(t)$ maintains its previous value when the condition (\ref{6:eq:condition}) is not satisfied:
\begin{equation}
\label{6:eq:price}
P(t)=\begin{cases}
	({\rm max}\{B_i(t)\} + {\rm min}\{S_i(t)\})/2, & \text{if the condition (\ref{6:eq:condition}) is satisfied,}\\
	P(t-1), & \text{otherwise.}
\end{cases}
\end{equation}

The dealers change their prices in a unit time by the following deterministic rule:
\begin{equation}
\label{6:eq:priceevo}
B_i(t+1)=B_i(t)+a_i(t)+c\Delta P(t),
\end{equation}
where $a_i(t)$ denotes the $i$th dealer\rq s characteristic movement in the price at time $t$, $\Delta P(t)$ is the difference between the price at time $t$ and the price at the time when the previous trade was done, and $c(>0)$ is a constant which specifies dealers\rq\ response to the market price change, and is common to all of the dealers in the market. 
The absolute value of a dealer\rq s characteristic movement $a_i(t)$ is given by a uniform random number in the range $[0,\alpha)$ and is fixed throughout the time.
The sign of $a_i$ is positive when the $i$th dealer is a buyer and is negative when the dealer is a seller.
The buyer (seller) dealers move their prices up (down) until the condition (\ref{6:eq:condition}) is satisfied.
Once the transaction takes place, the buyer of the transaction becomes a seller and the seller of the transaction becomes a buyer; in other words, the buyer dealer changes the sign of $a_i$ from positive to negative and the seller dealer changes it from negative to positive:
\begin{equation}
\label{6:eq:ai}
a_i(t+1)=
	\begin{cases}
		- a_i(t) & \text{for the buyer and the seller,}\\
		a_i(t) & \text{for other dealers.}
	\end{cases}
\end{equation}

The initial values of $\{B_i\}$ are given by uniform random numbers in the range $(-\Lambda,\Lambda)$. 
We thus simulate this model specifying the following four parameters: the number of dealers, $N$; the spread between the buying price and the selling price, $\Lambda$; the dealers\rq\ response to the market price change, $c$; and the average of dealers\rq\ characteristic movements in a unit time, $\alpha$. 

The ST model well reproduces the power-law behavior of the price change when the dealers\rq\ response to the market change $c>c^*$, where $c^*$ is a critical value to the power-law behavior.
The critical point depends on the other parameters; {\it e.g.} $c^*\simeq 0.25$ for $N=100$, $\Lambda=1.0$ and $\alpha=0.01$ \cite{6:sato}. 

\subsection{Microscopic model of triangular arbitrage: interacting two systems of the ST model}
\label{6:micromodel2}
We now describe our microscopic model as a set of the ST models.
In order to reproduce effects of the triangular arbitrage, we prepare two systems of the ST model, the market $X$ and the market $Y$.
As is noted above, we prepare only two markets to reproduce the effect of the triangular arbitrage because we regard two of the three markets as one effective market.

The dealers in the markets $X$ and $Y$ change their bidding prices according to the ST model as follows:
\begin{align}
B_{i,X}(t+1)&=B_{i,X}(t)+a_{i,X}(t)+c\Delta P_X(t) \label{6:eq:evo1} {\rm\ and}\\
B_{i,Y}(t+1)&=B_{i,Y}(t)+a_{i,Y}(t)+c\Delta P_Y(t), \label{6:eq:evo2}
\end{align}
where $X$ and $Y$ denote the markets $X$ and $Y$, respectively. 
An intra-market transaction takes place when the condition
\begin{equation}
\label{6:eq:condition2}
{\rm max}\{B_{i,x}(t)\} \ge {\rm min}\{S_{i,x}(t)\},\quad x=X \mbox{\ or\ } Y
\end{equation}
is satisfied.
We assume that $\Lambda$ is common to the two markets.
The price $P_x(t)$ is renewed in analog to the ST model:
\begin{equation}
\label{6:eq:price2}
P_x(t)=\begin{cases}
	({\rm max}\{B_{i,x}(t)\} + {\rm min}\{S_{i,x}(t)\})/2, & \text{if the condition (\ref{6:eq:condition2}) is satisfied,}\\
	P_x(t-1), & \text{otherwise,}
\end{cases}
\end{equation}
where $x=X$ or $Y$.

\begin{figure}
		\centering
		\includegraphics[height=10cm]{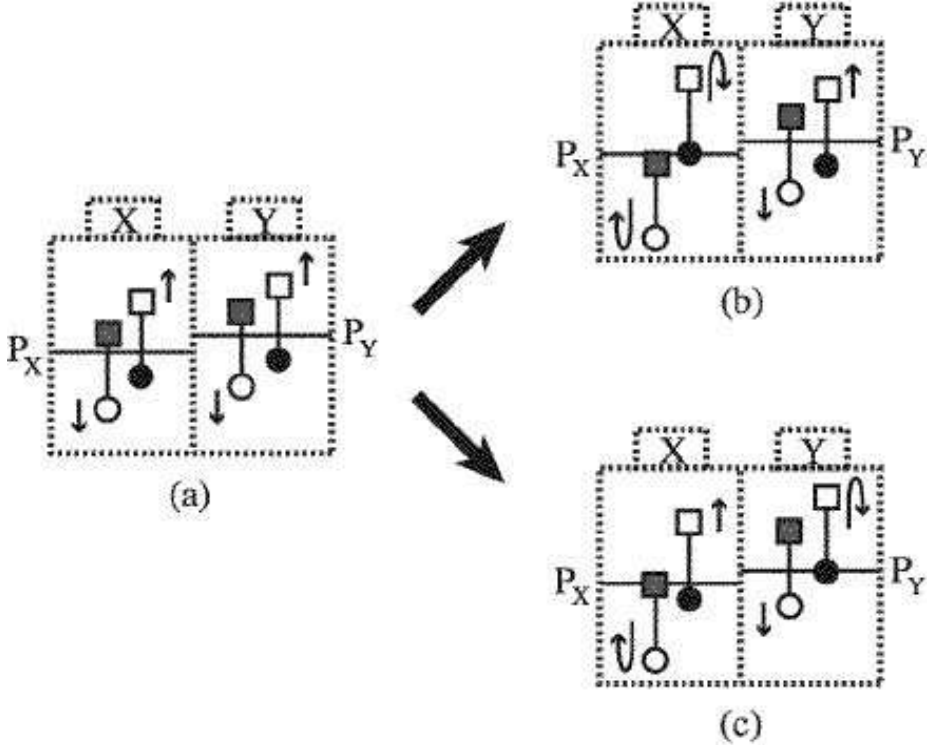}
	\caption{
	A schematic image of the transactions. 
	Only the best bidders in the markets are illustrated in order to simplify the image.	
	The circles and the squares denote the dealers\rq\ bidding price to buy and to sell.
	The filled circles denote the best bidding prices to buy in the markets, ${\rm max}\{B_{i,X}\}$ and ${\rm max}\{B_{i,Y}\}$, and the gray squares denote the best bidding prices to sell in the markets, ${\rm min}\{B_{i,X}\} + \Lambda$ and ${\rm min}\{B_{i,Y}\} + \Lambda$.
	In the case (a), any of the conditions (\ref{6:eq:condition2}), (\ref{6:eq:arb1}) and (\ref{6:eq:arb2}) are not satisfied.
	The buyers move their prices up, and the sellers move their prices down.
	In the case (b), the dealers in the market $X$ satisfy the condition (\ref{6:eq:condition2}); hence the intra-market transaction takes place.
	The price in the market $X$, $P_X$, is renewed, and the buyer and the seller of the transaction become a seller and a buyer, respectively.
	In the case (c), the seller in the market $X$ and the buyer in the market $Y$ satisfy the condition (\ref{6:eq:arb2}); hence the arbitrage transaction takes place.
	The price $P_X$ in the market $X$ becomes ${\rm min}\{B_{i,X}\} + \Lambda$, and the price $P_Y$ in the market $Y$ becomes ${\rm max}\{B_{i,Y}\}$.
	The buyer and the seller of the transaction become a seller and a buyer, respectively.
	The arbitrage transaction thus makes the interaction between the markets $X$ and $Y$.
	}
	    \label{6:fig:arb}
\end{figure}

We here add a new {\it inter}-market transaction rule which makes the systems interact.
The arbitrage transaction can take place when one of the conditions 
\begin{align}
		\nu_X\equiv &\ {\rm max}\{B_{i,X}(t)\}  - ({\rm min}\{B_{i,Y}(t)\} + \Lambda) \ge 0 
		\label{6:eq:arb1}\\ 
		\nu_Y\equiv &\ {\rm max}\{B_{i,Y}(t)\}	-  ({\rm min}\{B_{i,X}(t)\} + \Lambda ) \ge 0
		\label{6:eq:arb2}
\end{align}
is satisfied (see Fig.~\ref{6:fig:arb}). 
When the conditions (\ref{6:eq:condition2}) and (\ref{6:eq:arb1}) or (\ref{6:eq:arb2}) are both satisfied simultaneously, the condition (\ref{6:eq:condition2}) precedes.

Note that the arbitrage conditions  $\nu_X\ge 0$ and $\nu_Y\geq 0$ in the microscopic model correspond to the arbitrage condition $\nu\geq 0$ in the actual market, where $\nu$ is defined by Eq.~(\ref{6:eq:defnu}). 
We assume that the dealers\rq\ bidding prices $\{B_i\}$ and $\{S_i\}$ correspond to the logarithm of the exchange rate, $\ln r_i$. 
Therefore, ${\rm max}\{B_{i,X}\}$ may be equivalent to $-\ln(\mbox{yen-dollar ask})$ while ${\rm min}\{S_{i,Y}\}$ may be equivalent to $\ln(\mbox{dollar-euro ask}) - \ln(\mbox{yen-euro bid})$, and hence $\nu_X$ may be equivalent to $\nu$.
More precisely, the direction of the arbitrage transaction determines which of the quantities, $\nu_X$ or $\nu_Y$, corresponds to the logarithm rate product $\nu$.
There are two directions of the triangular arbitrage transaction.
The definition (\ref{6:eq:defnu}) specifically has the direction of Japanese yen to US dollar to euro to Japanese yen.
As is mentioned in Section \ref{6:macromodel}, we can define another logarithm rate product $\nu' $ in the actual market which has the opposite direction to $\nu$, Japanese yen to euro to US dollar to Japanese yen. 
Hence, if the logarithm rate product $\nu$ in the actual market corresponds to $\nu_X$ in Eq.~(\ref{6:eq:arb1}), $\nu\rq$ corresponds to $-\nu_Y$ in Eq.~(\ref{6:eq:arb2}).

The procedures of the simulation of the microscopic model are as follows (Fig.~\ref{6:fig:arb}):
\begin{enumerate}
\item Prepare two systems of the ST model, the market $X$ and the market $Y$, as described in Section \ref{6:STmodel}. The parameters are common to the two systems.
\item Check the condition (\ref{6:eq:condition2}) and renew the prices by Eq.~(\ref{6:eq:price2}).
If the condition (\ref{6:eq:condition2}) is satisfied, skip the step 3 and proceed to the step 4. Otherwise, proceed to the step 3.
\item Check the arbitrage conditions (\ref{6:eq:arb1}) and (\ref{6:eq:arb2}). 
If the condition (\ref{6:eq:arb1}) is satisfied, renew the prices $P_X(t)$ and $P_Y(t)$ to ${\rm max}\{B_{i,X}(t)\}$ and ${\rm min}\{B_{i,Y}(t)\} + \Lambda$, respectively.
If the condition (\ref{6:eq:arb2}) is satisfied, renew the prices $P_X(t)$ and $P_Y(t)$ to ${\rm min}\{B_{i,X}(t)\} + \Lambda$ and ${\rm max}\{B_{i,Y}(t)\}$, respectively.
If both of the conditions in (\ref{6:eq:arb1}) and (\ref{6:eq:arb2}) are satisfied, choose one of them with the probability of $50\%$ and carry out the arbitrage transaction as described just above.
If the arbitrage transaction takes place, proceed to the step 4; otherwise skip the step 4 and proceed to the step 5.
\item Calculate the difference between the new prices and the previous prices, $\Delta P_X(t) = P_X(t) -P_X(t-1)$ and $\Delta P_Y(t) = P_Y(t) -P_Y(t-1)$, and use them in the Eqs.~(\ref{6:eq:evo1}) and (\ref{6:eq:evo2}), respectively.
Change the buyer and the seller of the transaction to a seller and a buyer, respectively.
In  other words, change the signs of $a_{i,X}$ and $a_{i,Y}$ of the dealers who transacted.
\item If any of the conditions (\ref{6:eq:condition2}), (\ref{6:eq:arb1}) and (\ref{6:eq:arb2}) are not satisfied, maintain the previous prices, $P_X(t)=P_X(t-1)$ and $P_Y(t)=P_Y(t-1)$, as well as the previous price differences, $\Delta P_X(t) = \Delta P_X(t-1)$ and $\Delta P_Y(t) = \Delta P_Y(t-1)$.
\item Change the dealers\rq\ bidding prices following Eqs.~(\ref{6:eq:evo1}) and (\ref{6:eq:evo2}).
\item Repeat the steps from 2 to 6.
\end{enumerate}

\begin{figure}
	\begin{minipage}{6cm}
	\begin{center}
 	\includegraphics[height=5cm]{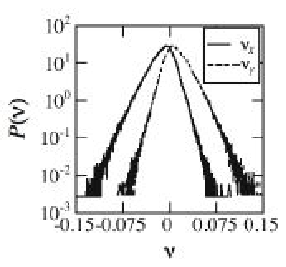}\\
	 \quad\quad
	(a)
	\end{center}
 	\end{minipage}
 	\begin{minipage}{8cm}
	 \begin{center}
	 \hspace{3cm}
 	\includegraphics[height=5cm]{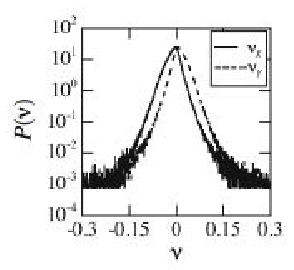}\\
	\quad
	 (b)
	 \end{center}
 	\end{minipage}
	\caption{The distributions of $\nu_X$ and $\nu_Y$. 
	The parameters are fixed to $N=100$, $\alpha=0.01$, $\Lambda=1.0$ and (a) $c=0.0$, (b) $c=0.3$, and are common to the market $X$ and the market $Y$. 
	The solid line denotes $\nu_X$ and the dashed line denotes $\nu_Y$ in each graph.
	}
	\label{6:fig:diff}
\end{figure}

\begin{figure}
	\begin{minipage}{6.5cm}
\begin{center}
\includegraphics[height=6cm]{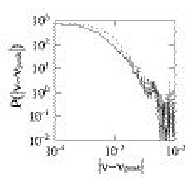}\\
(a)
\end{center}
 	\end{minipage}
 \quad
 	\begin{minipage}{6.5cm}
\begin{center}
\includegraphics[height=6cm]{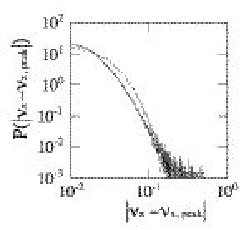}\\
(b)
\end{center}
 	\end{minipage}
\caption{The probability distribution of the difference between the logarithm rate product and its average for (a) the actual data, $|\nu-\nu_{\mathrm{peak}}|$ and for (b) the data of the microscopic model, $|\nu_x-\nu_{x, {\mathrm{peak}}}|$ for $c=0.3$, where $\nu_{\mathrm{peak}}$ and $\nu_{x, {\mathrm{peak}}}$ denote the peak position of respective data.
			The solid lines represent the part of the positive difference and the dotted lines represent the part of the negative difference, in the both graphs.
				We can see that the probability distribution of the logarithm rate product $\nu$ has a skewness around its average, and the microscopic model qualitatively reproduces it well.}
\label{fig:skew}
\end{figure}

The quantities $\nu_X$ and $\nu_Y$ are shown in Fig.~\ref{6:fig:diff}. 
(Following Ref.~\cite{6:sato}, we set the parameters common to the two markets $X$ and $Y$: $N=100$, $\alpha=0.01$ and $\Lambda=1.0$.)
In Fig.~\ref{6:fig:diff}(b) for $c=0.3$, the fat-tail behavior of the price difference $\nu_X$ is consistent with the actual data as well as with the macroscopic model in Fig.~\ref{6:fig:compfil}.
Furthermore, $\nu_X$ reproduces the skewness of the actual data, which cannot be reproduced by the macroscopic model (Fig.~\ref{fig:skew}).
Note that the skewness of $\nu_Y$ is consistent with the behavior of $\nu\rq$.

\section{The microscopic parameters and the macroscopic spring constant}
\label{6:micromacro}
In this Section, we discuss the relation between the macroscopic model and the microscopic model through the interaction strength, or the spring constant $k$.

\begin{figure}
\begin{minipage}{5.5cm}
 \includegraphics[height=5cm]{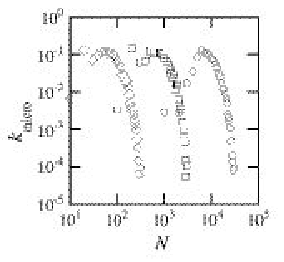}\\
 \begin{center}
 (a)
 \end{center}
 \end{minipage}
 \quad\quad
 \begin{minipage}{5.5cm}
 \includegraphics[height=5cm]{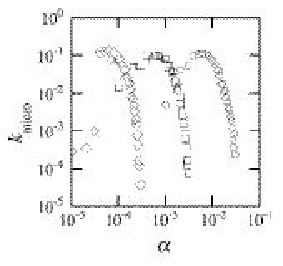}\\
 \begin{center}
 (b)
 \end{center}
 \end{minipage}
\caption{The spring constant $k_{\rm micro}$ as a function of parameters. 
			  The panel (a) shows the dependence on $N$. 
			  The other parameters are fixed to $\Lambda=1.0$ and $\alpha=0.0001, 0.001$ and $0.01$ for the circles, the squares and the diamonds, respectively, and $c=0.3$ for all the plots.
			  The panel (b) shows the dependence on $\alpha$. The other parameters are fixed to $\Lambda=1.0$ and $N=100, 1000$ and $10000$ for the circles, the squares and the diamonds, respectively, and $c=0.3$ for all the plots.}
\label{6:fig:barabara}
\end{figure}

In the microscopic model, we define the spring constant $k_{\rm micro}$, which corresponds to the spring constant $k$ of the macroscopic model, as follows:
\begin{equation}
\label{6:eq:microspring}
k_{\rm micro}\equiv \frac{1}{2}\left(1-\frac{\langle\nu_X(t+1)\nu_X(t)\rangle-\langle\nu_X(t)\rangle^2}{\langle\nu_X(t)^2\rangle-\langle\nu_X(t)\rangle^2
}\right).
\end{equation}
Figure~\ref{6:fig:barabara} shows the estimate (\ref{6:eq:microspring}) as a function of several parameters.

Remember that, in the macroscopic model, the spring constant $k$ depends on the time step $T$ (see Fig.~\ref{6:fig:aT}).
The spring constant of the microscopic model $k_{\rm micro}$ also depends on a time scale as follows.
The time scale of the ST model may be given by the following combination of the parameters \cite{6:sato}:
\begin{equation}
\label{6:eq:n}
\langle n\rangle \simeq\frac{3 \Lambda}{N\alpha},
\end{equation}
where $n$ denotes the interval between two consecutive trades.
Hence, the inverse of Eq.~(\ref{6:eq:n}), 
\begin{equation}
f\equiv 1/\langle n\rangle \simeq \frac{N\alpha}{3\Lambda}, 
\label{6:eq:deff}
\end{equation}
is the frequency of the trades.

Although there are four parameters $N$, $\alpha$, $\Lambda$ and $c$, we change only three parameters $N$, $\alpha$, and $c$ and set $\Lambda=1.0$, because only the ratios $N/\Lambda$ and $\alpha/\Lambda$ are relevant in this system.
The ratio $N/\Lambda$ controls the density of the dealers and $\alpha/\Lambda$ controls the speed of the dealers\rq\ motion on average. 
Hence, we set $\Lambda=1.0$ and change the other parameters hereafter.

\begin{figure}
	\begin{minipage}{7.5cm}
 		\includegraphics[height=7.5cm]{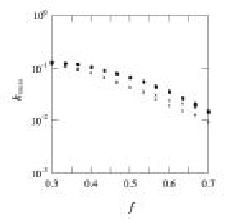}\\
		 \begin{center}
		 (a)
		 \end{center}
 	\end{minipage}
 	\begin{minipage}{7.5cm}
 		\includegraphics[height=7.5cm]{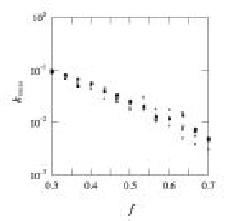}\\
		 \begin{center}
		 (b)
		 \end{center}
 	\end{minipage}
\caption{The scaling plot of the spring constant $k_{\rm micro}$ as a function of the trade frequency $f=N\alpha/3\Lambda$. 
			The vertical axes are displayed in the logarithmic scale. The dealers\rq\ response to the price change, $c$, is fixed to $0.0$ in (a) and $0.3$ in (b). 
			We fix $\alpha=0.0001, 0.001$ and $0.01$ and change $N$ (open circles, squares, and diamonds, respectively) and $N=100, 1000$ and $10000$ and change $\alpha$ (crosses, filled circles and triangles, respectively), while $\Lambda$ is fixed to $1$.  
			Note that all points collapse onto a single curve.
			The spring constant $k_{\rm micro}$ is scaled by $f$, and decays exponentially in both of the plots (a) and (b). 
			} 
\label{6:fig:scalingplot}
\end{figure}

We plot the spring constant $k_{\rm micro}$ as a function of the trade frequency $f\equiv N\alpha/3\Lambda$ in Fig.~\ref{6:fig:scalingplot}.
The plots show that the spring constant $k_{\rm micro}(N, \alpha, \Lambda)$ can be scaled by the trade frequency $f$ well.

\begin{figure}
\includegraphics[height=5cm]{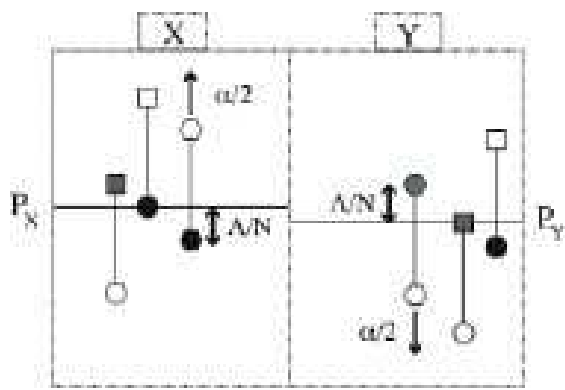}
\caption{A schematic image of the second best bidders\rq\ motion.
			The circles and the squares denote the dealers\rq\ bidding price to buy and to sell.
			The filled circles and the gray squares represent the best bidding prices to buy and sell, respectively.
			The hexagons denote the second best bidding prices to buy (the filled one) and to sell (the gray one).
			}
\label{6:fig:darb}
\end{figure}

In order to determine a reasonable range of the parameters, let us consider the situation in Fig.~\ref{6:fig:darb}, where the arbitrage transaction is about to take place.
At the moment, the positions of the second best bidders (hexagons) in the markets $X$ and $Y$ are, on average, $\Lambda/N$ away from the prices transacted, $P_X$ and $P_Y$.
In the next step, the second best bidders in the markets $X$ and $Y$ will move by $\alpha/2$ on average toward to the prices $P_X$ and $P_Y$, respectively.
The next transaction will be carried out probably by the second best bidders.
For $\alpha/2 > \Lambda/N$, the prices of the transactions may move away from each other. 
The arbitrage transaction cannot bind the two prices of the markets $X$ and $Y$ enough and the two prices $P_X$ and $P_Y$ fluctuate rather freely.
It is not a realistic situation.
Therefore, the condition 
\begin{equation}
f = \frac{N\alpha}{3\Lambda} \leq \frac{2}{3}
\label{6:eq:2/3}
\end{equation}
should be satisfied for the real market to be reproduced. 
On the other side, the simulation data have too large errors in the region $f<1/3$ because the transaction rarely occurs.
We hence use the data in the region $1/3\leq f\leq2/3$ hereafter.

\begin{figure}
\includegraphics[height=6cm]{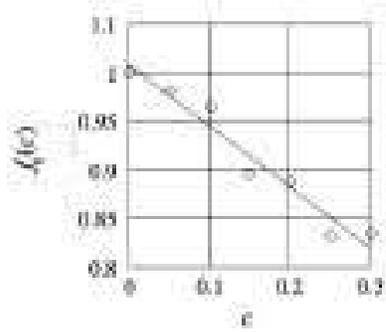}
\caption{The dependence of $f_0(c)$ on $c$, estimated by fitting the data in Fig.~\ref{6:fig:scalingplot} as well as the same plots for different values of $c$.}
\label{6:fig:beta}
\end{figure}

The spring constant $k_{\rm micro}$ decays exponentially in the range $1/3\leq f\leq 2/3$ in both of the plots (a) and (b) of Fig.~\ref{6:fig:scalingplot}, having different slopes.
Hence we assume that the spring constant decays as 
\begin{equation}
\label{6:eq:exp} 
k_{\rm micro}\propto e^{-f/f_0(c)},
\end{equation}  
where $f_0(c)$ denotes the characteristic frequency dependent on $c$.
The estimates of the characteristic frequency $f_0(c)$ are shown in Fig.~\ref{6:fig:beta} as a function of $c$.
The characteristic frequency $f_0(c)$ thus estimated decays linearly with $c$.
The reason why $f_0(c)$ behaves so is an open problem.

\begin{figure}
\includegraphics[height=5cm]{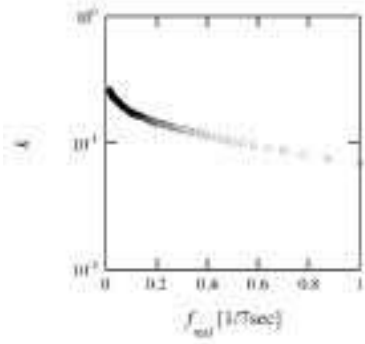}
\caption{We plotted the same data as in Fig. \ref{6:fig:aT}, but the horizontal axis here is the trade frequency scaled by the realistic time scale of the trades, $T_{\mathrm{real}}=7$[sec]. }
\label{6:fig:k_real}
\end{figure}

In Fig.~\ref{6:fig:k_real}, we plot the same data as in Fig. \ref{6:fig:aT}, but by making the horizontal axis the trade frequency $f_{\mathrm{real}}$.
In order to compare it with Fig. \ref{6:fig:scalingplot} quantitatively, we used the time scale $T_{\mathrm{real}}=7$[sec]; the interval between two consecutive trades in the actual foreign exchange market is, on average, known to be about $7$[sec] \cite{6:ohira}.
The spring constant in the actual market $k$ decays exponentially with the trade frequency $f_{\mathrm{real}}$, which is consistent with that of the microscopic model shown in Fig. \ref{6:fig:scalingplot}.
The real characteristic frequency in Fig. \ref{6:fig:k_real}, however, is quite different from that of the microscopic model plotted in Fig. \ref{6:fig:scalingplot}.
This is also an open problem.

\begin{figure}[htbp]
	\begin{minipage}{6cm}
		 \begin{center}
 		\includegraphics[height=6cm]{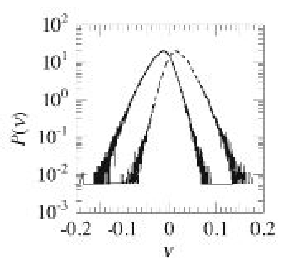}\\
		 (a)
		 \end{center}
 	\end{minipage}
 \quad
 	\begin{minipage}{6cm}
	 \begin{center}
 		\includegraphics[height=6cm]{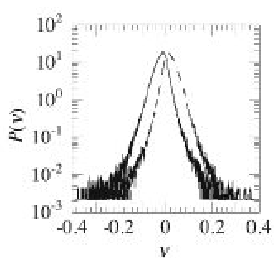}\\
		 (b)
	\end{center}
 	\end{minipage}
\caption{The distributions of the logarithm rate products $\nu_X$ and $\nu_Y$ for the microscopic model consisting of three ST models. 
				The solid lines represent $\nu_X$ and the dashed lines represent $\nu_Y$.
				The parameters are common to the three markets as $N=100$, $\Lambda=1.0$, $\alpha=0.01$ and (a) $c=0.0$, and (b) $c=0.1$. 
				In (b) the model well reproduces the fat-tail behavior of the actual data.
				}
\label{fig:3markets}
\end{figure}

We have so far discussed the microscopic model consisting of two ST models.
The model well describes the actual data qualitatively.
We here note that we can reproduce the actual data by preparing straightforwardly a microscopic model consisting of three ST models (see Fig. ~\ref{fig:3markets}).

\section{Summary}\label{6:sum}
We first showed in Ref.~\cite{6:tat1} that triangular arbitrage opportunities exist in the foreign exchange market. 
The rate product $\mu$ fluctuates around its average.
We then introduced in Ref.~\cite{6:tat1} the macroscopic model including the interaction caused by the triangular arbitrage transaction.
Inspite of its simpleness, the maacroscopic model reproduced the actual behavior of the logarithm rate product $\nu$ well.
We finally introduced here the microscopic model, which consists of two systems of the ST model.
The microscopic model also reproduced the actual behavior of the logarithm rate product $\nu$ well.
The microscopic model can describe more details than the macroscopic model, in particular, the skewness of the distribution of the logarithm rate product $\nu$.
We then explored the relation between the spring constant of the macroscopic model and the parameters in the microscopic model.
The spring constant of the microscopic model $k_{\mathrm{micro}}$ can be scaled by the trade frequency $f$, and it decays exponentially with $f$, which is consistent with the spring constant of the actual market $k$.


\end{document}